\documentclass[aps,prl,twocolumn,superscriptaddress,floatfix]{revtex4-2}

\usepackage{graphicx}
\usepackage{color}
\usepackage[version=4]{mhchem} 
\usepackage[caption=false]{subfig} 
\usepackage{lineno}
\usepackage{grffile}

\begin{document}
\sloppy

\title{Particle Detection Using Magnetic Avalanches in Single-Molecule Magnet Crystals}


\author{Bailey Kohn}
\email[]{b.pickard17@tamu.edu}
\affiliation{Department of Physics and Astronomy, Texas A\&M University, College Station, TX, USA}
\author{Hao Chen}
\email[]{chenhao\_fd@fudan.edu.cn}
\affiliation{Department of Physics and Astronomy, Texas A\&M University, College Station, TX, USA}
\affiliation{Institute of Modern Physics, Fudan University, Shanghai 200433, China}
\author{Rupak Mahapatra}
\author{Glenn Agnolet}
\author{Ivan Borzenets}
\affiliation{Department of Physics and Astronomy, Texas A\&M University, College Station, TX, USA}
\author{Philip C. Bunting}
\affiliation{Departments of Chemistry, Chemical and Biomolecular Engineering, and Materials Science and Engineering, University of California, Berkeley, CA, USA}
\author{Jeffrey R. Long}
\affiliation{Departments of Chemistry, Chemical and Biomolecular Engineering, and Materials Science and Engineering, University of California, Berkeley, CA, USA }
\affiliation{Materials Sciences Division, Lawrence Berkeley National Laboratory, Berkeley, CA, USA.}
\author{Minjie Lu}
\affiliation{Department of Physics and Astronomy, Texas A\&M University, College Station, TX, USA}
\author{Tom Melia}
\affiliation{Kavli IPMU (WPI), UTIAS, The University of Tokyo, Kashiwa, Chiba 277-8583, Japan} 
\author{Michael Nippe}
\affiliation{Department of Chemistry, Texas A\&M University, College Station, TX, USA}
\author{Lok Raj Pant}
\affiliation{Department of Physics and Astronomy, Texas A\&M University, College Station, TX, USA}
\author{Surjeet Rajendran}
\affiliation{Department of Physics \& Astronomy, The Johns Hopkins University, Baltimore, Maryland 21218}
\author{Anna Schmautz}
\affiliation{Department of Chemistry, Texas A\&M University, College Station, TX, USA}
\author{Amis Sharma}
\affiliation{Department of Physics and Astronomy, Texas A\&M University, College Station, TX, USA}


\date{\today}

\begin{abstract}

The detection of a single quantum of energy with high efficiency and a low false positive rate is of considerable scientific interest, from serving as single quantum sensors of optical and infra-red photons to enabling the direct detection of low-mass dark matter. We confirm our initial experimental demonstration of magnetic avalanches induced by scattering of quanta in single-molecule magnet (SMM) crystals made of Mn$_{12}$-acetate, establishing the use of SMMs as particle detectors for the first time. Although the current setup has an energy threshold in the MeV regime, our results motivate the exploration of a wide variety of SMMs whose properties could allow for detection of sub-eV energy depositions.

\end{abstract}


\maketitle

\begin{figure}[!htbp]
    \centering
    \includegraphics[width=\linewidth]{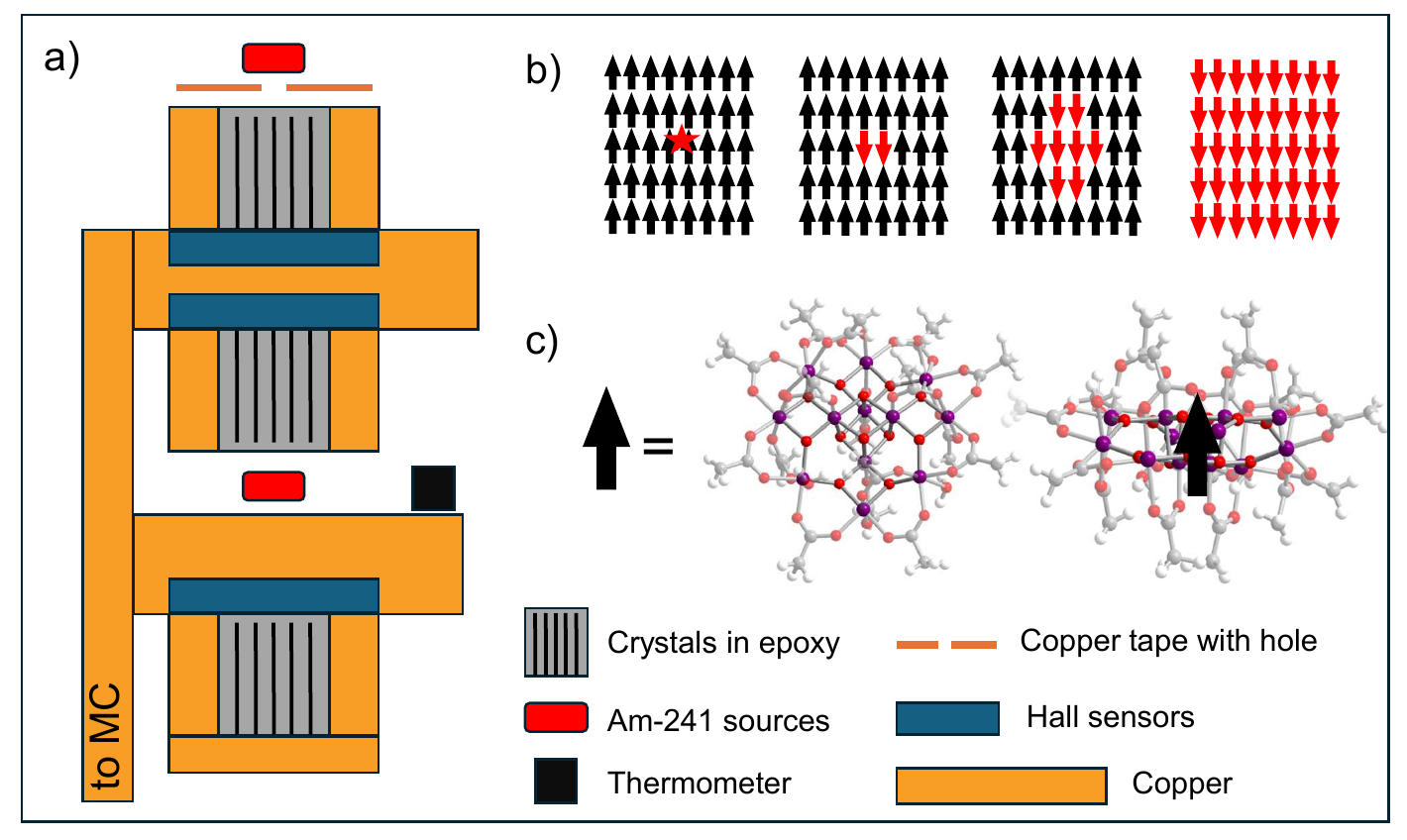}
    \caption{a) Schematic view of the experimental setup. Crystals of Mn$_{12}$-acetate were mounted using silver epoxy in three sample holders. The sample holders were each thermally connected to the mixing chamber and equipped with an independent hall sensor. $^{241}$Am $\alpha$ sources were situated near two of the sample holders, directly facing the crystals within. One source was collimated with copper tape (top) and the other was left open (middle). The crystals inside the final sample holder (bottom) were fully shielded from the sources. A resistive thermometer was placed nearby and thermally connected to the system. b) Conceptual illustration of a single-molecule magnet-based particle detector at the molecular level. An interaction deposits some energy at a crystal site, the deposited energy locally heats the crystal, causing some of the spins to relax, releasing their Zeeman energy. The released energy further heats the crystal locally, causing nearby spins to also relax. The avalanche process continues until the whole crystal relaxes, with a measurable change in the crystal magnetization. c) Chemical structure of a single molecule. (left) Crystal structure of Mn$_{12}$-acetate, where eight outer S = 2 Mn(III) ions are antiferromagnetically coupled to four central-cubane-based S = 3/2 Mn(IV) ions to give an S = 10 ground state. Purple, red, gray, and white spheres represent Mn, O, C, and H, respectively. Acetate ligands are faded to aid visualization of the inorganic Mn$_{12}$ core. (right) View of Mn$_{12}$-acetate with the molecular magnetic easy axis—denoted by faded black arrow—aligned with the external polarizing magnetic field.}
    \label{fig:diagram}
\end{figure}

It is scientifically challenging to develop sensors that can detect energy depositions in the sub-eV range with high efficiency and low false positive (or dark count) rates. Sensors with this capability can be used to count single quanta of infra-red photons, a technical feat that has broad applications to many fields \cite{Komiyama2000, Lita:s, Eisaman2011}, including quantum computing \cite{Ladd2010, Knill2001}. Such single quantum sensors \cite{HVeV2019, Rong2018} may also open a path toward the detection of the scattering of sub-GeV dark matter and the absorption of sub-eV dark matter such as dark photons \cite{Bunting2017}. This is a theoretically well motivated region of dark matter parameter space \cite{PhysRevLett.121.101801, Battaglieri:2017aum, PhysRevD.93.103520} that has prompted several collaborations \cite{TESSERACT_recent_results, CRESST_first_results, CRESST_recent_results, EDELWEISS, SuperCDMS, DAMIC-M, SENSEI}. The detection of small energies can be accomplished through the use of an amplification technique that magnifies the effect of the initial energy deposition. This can be implemented through applied voltage \cite{HVeV2019, Hybrid_Detector} or internal amplification \cite{Chesi2019, Knopfmacher2014, Sorgenfrei2011}. One such example is that of bubble chamber threshold detectors where liquid is held in a superheated state such that particle scattering will cause bubbles to form ~\cite{PhysRevD.100.022001}. Recently, an analogous detector has been proposed by Bunting et al. using the properties of single-molecule magnets (SMMs) \cite{Bunting2017}.
\begin{figure}[!htbp]
    \centering
    \includegraphics[width=\linewidth]{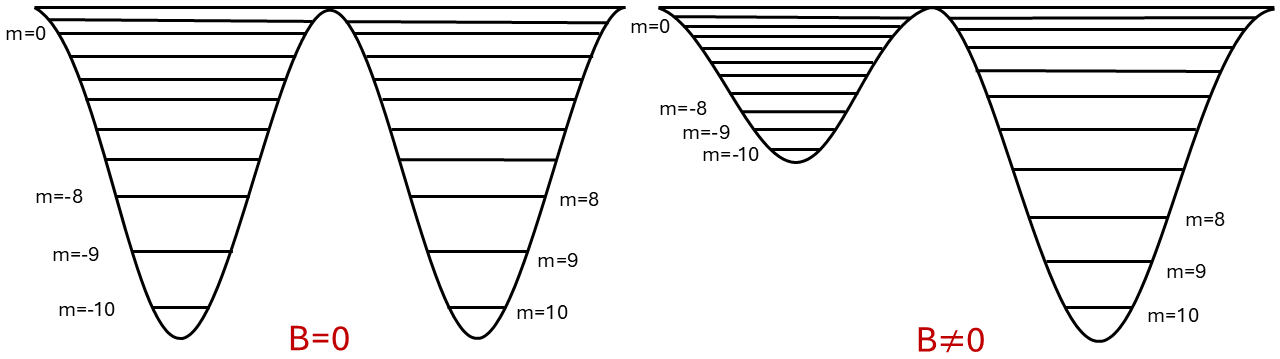}
    \caption{The magnetic spin potential of Mn$_{12}$-ac in zero magnetic field (left) and in a non-zero magnetic field (right). With no external field present, the potential is doubly degenerate, whereas if a field is applied the potential has a ground state and a meta-stable state.}
    \label{fig:potentials}
\end{figure}
\\\indent
First discovered just over 30 years ago \cite{1993Natur.365..141S}, SMMs exhibit magnetic bistability and a barrier to magnetization reorientation, which can lead to phenomena such as magnetic hysteresis at low temperatures. As such, these molecules have garnered substantial interest for potential applications including spin-based electronics ~\cite{Bogani2008} and quantum computing ~\cite{Leuenberger2001}. Application of a static magnetic field lifts the degeneracy of the molecular magnetic ground state, giving rise to a metastable state in which molecules can remain trapped for several months at cryogenic temperatures ~\cite{Barra_Fe8_original}. It is known that under certain circumstances, such as through a pulsed magnetic field, local heating, surface acoustic waves, or a resonance field, a collection of single-molecule magnets in a crystal can undergo a reversal of magnetization, resulting in the release of their Zeeman energy. This release causes the process to repeat in neighboring molecules, resulting in a complete and rapid reversal of magnetization throughout the crystal; a process known as a magnetic avalanche \cite{Paulsen1995_first_avalanche, PhysRevLett.95.147201, PhysRevB.90.134405, PhysRevB.76.054410, PhysRevB.81.064437}. Magnetic avalanches have been studied in detail, triggered by mechanisms such as supplying energy via surface acoustic waves~\cite{PhysRevLett.95.217205, Macia_avalanche_acoustic_waves}, direct heating of one side of a crystal~\cite{McHugh2007_avalanche_by_heat, PhysRevLett.110.207203}, or sweeping the external magnetic field to directly alter the stability of the metastable state~\cite{PhysRevLett.95.147201}. In principle, a magnetic avalanche could also be triggered by small localized energy depositions as recently proposed~\cite{Bunting2017} (FIG. \ref{fig:diagram}.b)).

These molecular magnets are set apart from other candidate sensors in the literature because of their potential for a very low threshold of detection and unparalleled chemical tunability. In their paper, Bunting et al. calculated that, using a SMM with the right parameter, the detector threshold could be as low as 10 meV. This would make the detector sensitive to dark photons, a well motivated region of dark matter \cite{An_dark_photon_detection}. 
The energy threshold for SMMs is very dependent on the external field, creating the possibility for sensitive tunability. The spin potential barrier can be written as \cite{Bunting2017}
\begin{equation}
    \tilde{U}(B) = U -\frac{1}{2}\Delta E_{Zee}
\end{equation}
where $\Delta E_{Zee}$ is the Zeeman splitting energy defined by
\begin{equation}
    \Delta E_{Zee} = 2\mu_Bg_JJB
\end{equation}
where $\mu_B$ is the Bohr magnetron, $g_J$ is the Lande-g factor, $J$ is the spin of the molecule, and $B$ is the external magnetic field \cite{Bunting2017} (see FIG. \ref{fig:potentials}.). For an SMM with spin ($J=10$) $\Delta E_{Zee} \approx 0.7 meV$. As the field is increased, the barrier height that must be overcome for the spins to escape the metastable state decreases, thereby lowering the avalanche energy threshold. 

In this paper, we report confirmation of our first experimental demonstration~\cite{chenpaper, Chenthesis} of a magnetic avalanche triggered by $\alpha$ particle scattering in Mn$_{12}$O$_{12}$(O$_2$CCH$_3$)$_{16}$(H$_2$O)${_4}$ (Mn$_{12}$-ac, FIG. \ref{fig:diagram}.) ~\cite{1993Natur.365..141S} crystals and improved temperature data. Mn$_{12}$-ac is the first SMM discovered by Sessoli et al. in 1993 \cite{1993Natur.365..141S} and its properties have been extensively studied, so it is an ideal molecule for our proof-of-concept experiment. We synthesized the crystals according to the procedure outlined by Lis and North \cite{Mn12_synthesis, Norththesischemistry} and separated out the largest crystals which are roughly $2.5\times 0.5\times0.5$ mm$^3$. From the crystal structure of Mn$_{12}$-acetate \cite{Lis_Mn_structure}, we estimate the molecular density to be $~0.54\text{ molecules}/\text{nm}^3$, giving a detector mass of about $1$ mg. While our choice to use Mn$_{12}$-ac and an alpha source puts the energy threshold of our particular setup on the order of MeV, our results offer the first experimental proof-of-concept for the use of single-molecule magnets as potential quantum sensors. 


The particle detector setup, shown in FIG. \ref{fig:diagram}., features three different sample holders containing about 20 crystals each. The crystals are held in place with silver epoxy to act as a heat sink and keep them from being crushed by the force of the magnetic field. Since Mn$_{12}$-acetate grows along its easy axis, care was taken to align the crystals nearly parallel to the field; however, some were shifted off-axis when applying the epoxy leading to a smaller signal. The sample holders and their thermal links are made from high-conductivity, oxygen-free copper to ensure good heat transfer. Graphene hall sensors are placed near each sample holder and aligned with the easy axis of the crystals. A layer of Kapton tape lies between the hall sensors and the sample holders to provide a thermal barrier. The graphene hall sensors all have a sensitivity of $1300$ to $1600$ V/AT, giving a noise threshold on the order of $1$ Gauss when supplied with $100\ \mu$A of current. Americium 241 (Am-241) $\alpha$ sources are placed near two of the sample holders. Both are enclosed by containers having a circular opening $2.5$mm in diameter and sample 1 is covered with copper tape with a $1$mm diameter hole for further collimation. Both sources were measured with a Geiger counter and the open source was measured to have a strength of $\sim115$ CPM and the collimated source had a strength of $\sim100$ CPM. Care was taken to ensure that the silver epoxy did not cover the crystals closest to the sources to avoid blocking the $\alpha$ particles. The crystals in the last sample (hereafter called control) were fully covered with copper and epoxy to shield them from all $\alpha$ particles. All three samples, along with a resistive ruthenium oxide (RuOx) thermometer, were then placed symmetrically in a dilution fridge. They were mounted on a cold finger that is thermally connected to the mixing chamber stage (base 8 mK). The fridge is fitted with a 6 T superconducting magnet, and the cold finger extends into the center of the magnet.

All SMMs only display magnetic hysteresis below a certain temperature known as the blocking temperature which is defined in terms of the relaxation time. Many factors determine the relaxation time \cite{SMM_through_time_ZABALALEKUONA}, but near the blocking temperature, the relaxation time for Mn$_{12}$ can be modeled by the Arrhenius law:
\begin{equation}
    \tau = \tau_0 \text{exp}\left(\frac{k_BU-\frac{1}{2}\Delta E_{Zee}}{k_B T} \right)
\end{equation}
where $\tau_0$ and $U$ are intrinsic to the material \cite{Barra_Fe8_original}. Generally, the blocking temperature is given as the temperature in which molecules relax in 100s at zero field, however it can be reported for other timescales if more useful. For a low-count dark matter experiment, a longer timescale is ideal. For Mn$_{12}$-acetate, $U=61$ K and $\tau_0 = 2.1\times10^{-7}$ s ~\cite{1993Natur.365..141S} giving a blocking temperature of $\sim 2$ K for a timescale of 2 months in zero field \cite{Bunting2017, Sessoli1993}. We first verified this by magnetizing the crystals according to the method outlined in the next paragraph and then removing the external field. The temperature was then incrementally set to specific values using the heater on the mixing chamber and held steady at each value for $\sim 10$ minutes. The highest temperature at which the crystals did not clearly lose their magnetization during those 10 minutes was indeed found to be $2.0$ K.


Measurements were taken at two different temperatures: $\sim1.7$ K and $\sim 800$ mK. Since both are below the blocking temperature,  avalanches are expected at both values. Before collecting each set of data, the crystals were first magnetized by turning the external magnetic field to $\pm1.0$ T for the high temperature measurement or $\pm2.0$ T for the low temperature measurements, and then warming the crystals above their blocking temperature. To warm in the high temperature range, the built-in mixing chamber stage heater was set to $4$ K. Setting this heater to $>2$ K would destabilize the dilution process in the low temperature range, so we applied DC bias current through the hall sensors to locally heat the samples. To ensure the samples were warmed above their blocking temperature, we waited until the thermometer secured to the cold finger read above $2$ K for at least $30$ s. For both temperatures, the mixing chamber was then allowed to cool to operating temperature before the field was slowly brought ($\sim 0.1$ T/min) to $0$ T so that the crystals would remain magnetized in the orientation of the original applied field.


\begin{figure}[!htbp]
    \centering
    \includegraphics[width=\linewidth]{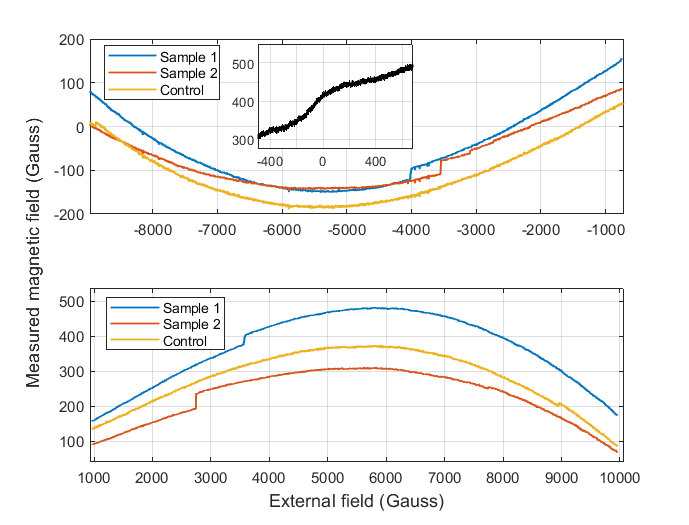}
    \caption{$\alpha$ particle induced avalanches observed in \ce{Mn12}-acetate during a constantly ramping field. The y-axis is hall sensor data subtracted from the applied external magnetic field. (top) Avalanches observed using a continuous reverse field ramp from $0$ T to $-1$ T at $1.7$ K. The two avalanches occur in a similar external field and the change in magnetization occurs in less than 1 s. There is no signal in the control sample at the time of either avalanche. Small avalanches can be seen in multiple samples, including the control, above $7.5$ T which is due only to the external field. (insert) Example of quantum tunneling around $0$ T external field seen in hysteresis loop. The slope is much shallower than a typical avalanche. (bottom) Avalanches observed using a continuous reverse field ramp from $0$ T to $+1$ T at $800$ mK. This is still the reverse field as the crystals were initialized in a negative field for this run. Similar results were observed as in the high temperature measurement.}
    \label{fig:rampingfield}
\end{figure}

\begin{figure}
    \centering
    \includegraphics[width=\linewidth]{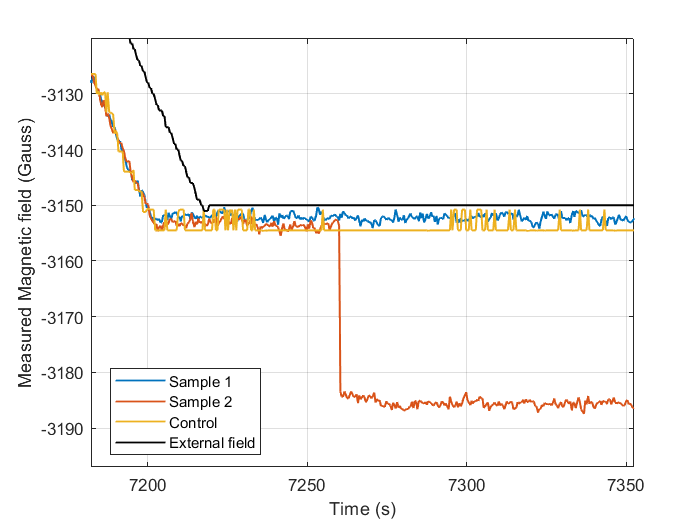}
    \caption{Example of an $\alpha$ particle induced avalanche observed at constant external field in \ce{Mn12}-acetate. An avalanche was observed only in sample 2 at the time of the avalanche. The slow increase in the hall sensor signal in all samples is due to the external field ramp up, whereas the sharp change is due to a magnetic avalanche that occurs in less than 1 s. The sensitivity in the control's lock-in amplifier was set lower than the others during this measurement, leading to the quantized noise. This noise is still well within the size of expected avalanche so it can be safely assumed that no avalanche occurred in the control at this field.}
\label{fig:constantfield}
\end{figure}

The threshold for triggering an avalanche in Mn$_{12}$-acetate depends on the external field and is difficult to estimate under our conditions, largely because the thermal conductivity of the crystals in the epoxy is unknown. As such, we first searched for avalanches by continuously ramping the field opposite of crystal magnetization at a rate of $0.025$ T/min. Three of these scans were performed, the first at high temperature and the rest at low temperature. FIG. \ref{fig:rampingfield}. shows results from runs 1 and 3. Runs 1 and 2 were magnetized in a positive external field and data was taken as the field was ramped from $0$ T to $-1$ T. Run 3 was magnetized in a negative external field and data was collected as the field was ramped from $0$ T to $+1$ T at the same rate. We assume a linear fit when converting the hall sensor voltage data to Gauss. This does not take into account the magnetic hysteresis that arises over our field sweep, so the result is a parabolic shape for the hall sensor data. In each scan, an avalanche was observed in both source samples in the $0.3$ T to $0.4$ T reverse field range. Note that the control sample did not experience an avalanche in the same range. Each jump ranged from $20$ Gauss to $50$ Gauss, which is the same order of magnitude as is expected for one crystal to completely avalanche assuming perfect alignment ~\cite{Chenthesis} (smaller jumps correspond to misaligned crystals). Each jump was also very sharp ($<0.5$ s, see FIG. \ref{fig:constantfield}.), a clear difference from the slow rise indicative of magnetic quantum tunneling at this temperature ~\cite{Friedman_2010} (FIG. \ref{fig:rampingfield}. inset). Small jumps of $\sim10$ Gauss were observed in multiple samples, including the control, in the range above $0.75$ T. The energy barrier becomes significantly lower in high external fields (eq 1), leading to the possibility of avalanches caused by lower energy background particles or temperature fluctuations. Furthermore, avalanches have been shown to be caused solely by a high reverse field \cite{Chenthesis, PhysRevLett.95.147201}, so these avalanches, even in shielded crystals, are expected. 


After confirming we can create avalanches in a changing magnetic field, we tested for avalanches in a steady field. After magnetizing, the field was brought to a $0.25$ T reverse field (this is below the threshold of any observed avalanche). The field was then increased in steps of $50$ Gauss and held at each step for $2$ min to $4$ min until reaching $\sim 0.45$ T reverse field. Four separate runs were performed at $800$ mK; in three of them an avalanche was seen while the field was constant and in one an avalanche occurred during the ramping between steps. An example is shown in FIG. \ref{fig:constantfield}. Only sample 2 underwent magnetic avalanche during each of these tests. The source on sample 1 was collimated, giving rise to a lower probability of interaction. This is seen in the ramping field runs where sample 1 always underwent an avalanche after sample 2. Following this logic, an avalanche in sample 1 would have likely been seen had the duration of each step been longer.


In all runs that resulted in an avalanche, the magnetic field jumps were abrupt (occurring in less than 1 s) and comparable to those observed in the literature \cite{PhysRevLett.95.147201, McHugh2007_avalanche_by_heat} for magnetic avalanches triggered in single-molecule magnets by other means. Moreover, given the stability of the control sample over all runs, we can reasonably assume that other magnetic relaxation pathways, such as resonant quantum tunneling, do not play a role at these temperatures. We note that the probability that the observed transitions are due to some random phenomenon common to all the samples (e.g. vibrations, electrical glitches, unstable temperature, or background radiation) can be estimated (using standard Poisson statistics) from the control channel and is negligible. Together, these data strongly support the presence of magnetic avalanches in the Mn$_{12}$-acetate crystals induced by the absorption of $\alpha$ particles.

\begin{figure}
    \centering
    \includegraphics[width=\linewidth]{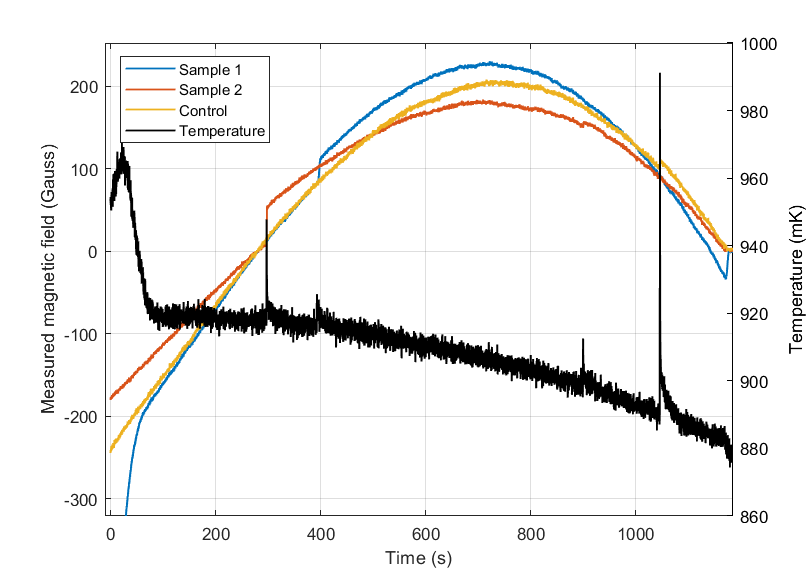}
    \caption{Ramping field magnetization vs time from Run 3 (bottom graph from FIG. \ref{fig:rampingfield}.) overlaid with temperature from the thermometer positioned on the cold finger. There is a temperature spike corresponding to the avalanches in all samples. The intensity of the temperature spikes is directly related to the external field. The downward trend in background temperature is due to the fridge continuing towards operating base temperature.}
\label{fig:temp}
\end{figure}

As one more test of this conclusion, we compare the hall sensor data with the temperature data from the RuOx thermometer on the cold finger. RuOx thermometers are insensitive to magnetic fields. Magnetization data from the third continuously ramping run (FIG. \ref{fig:rampingfield}. bottom) and temperature are plotted against time in FIG. \ref{fig:temp}. Since this run was performed around $800$ mK, the background temperature had low noise and any spikes could be clearly seen. For every avalanche, including those caused by high fields, there is a corresponding temperature spike. The temperature spike is expected and is due to the large amount of Zeeman energy released when the spins flip. From equation 2, we can calculate the Zeeman energy released by one molecule in a given magnetic field. Assuming every molecule is involved in an avalanche, the total released energy is
\begin{equation}
    E = 2\mu_Bg_JJB N = B \cdot 1.2\times10^{-4} \text{ J}.
\end{equation}
To calculate the change in heat, we use a basic heat equation:
\begin{equation}
    Q = cm\Delta T
\end{equation}
where c is specific heat and m is mass. We estimate the mass of the cold finger to be $0.01$kg and take $0.1$ kJ/K for the specific heat of copper at cryogenic temperatures. The temperature corresponding to the $\sim0.3$T avalanche in sample 2 is then calculated to be $\Delta T \approx 38$mK and the temperature change corresponding to the $\sim0.9$T avalanche in the control sample is $\Delta T \approx 95$mK. The small temperature spikes corresponding to avalanches in sample 2 are assumed to be explained by poor thermal contact. Because only one avalanche corresponds to each temperature spike, it can be safely assumed that the increase in temperature is the result, rather than the cause, of each avalanche.


The foregoing results definitively show that magnetic avalanches in crystals of Mn$_{12}$-acetate can be triggered by the absorption of elementary particles, with a threshold lower than 5.486 MeV for $\alpha$ particles at the values of magnetic field and temperature reported here. We note that the experimental energy detection threshold determined here for Mn$_{12}$-acetate is high compared to the 10 meV threshold desired for single infra-red photon or dark photon~\cite{An_dark_photon_detection, JUNGMAN1996195}. 
The expected threshold for an avalanche in single-molecule magnets depends on many parameters, such as molecular mass, thermal properties, and energy barrier \cite{Bunting2017}. Bunting et al. have shown in their paper that a threshold of $10$ meV can be achieved, for example, using an SMM with an energy barrier of $U=50$ K and a relaxation constant of $\tau_0=5\times 10^{-12}$ s, assuming typical values for all other parameters \cite{Bunting2017}. Luckily, SMMs with values in this range exist\cite{Mn-32_Manoli}.
Rigorous analysis will require determination of the specific heat capacities and thermal conductivities of candidate systems to better identify systems with appropriate threshold energies for detection of particles of various energies. 


The SMM-based magnetic avalanche detector concept offers unique capabilities as a detector for keV to MeV mass Dark Matter or CE$\nu$NS and meV scale “dark photon” experiments. First, the avalanches from an energy deposition can be contained. Instead of making one large crystal, the detector can be made from a number of smaller grains. Thus, deposited energy only causes the spins to relax in a single grain. With a sensitive magnetometer, spin flips of grain sizes $10$ to $100$ microns can be observed in a $1$ to $10$ cm sample. This containment could allow the detector to operate for long periods without the need to reset after every energy deposition. Second, we can create an array-based readout of SMM crystals to provide background veto based on crystal position, similar to what is done in other detector-based dark matter searches \cite{SuperCDMS}. Third, micro-fabrication techniques can be employed to place magnetometers, as well as thermal and other sensors, very close to the smaller SMM crystals, thus increasing resolution, reducing reset time, and providing veto capabilities from other crystals. Finally, depending on which SMM is used, we could probe either sub-GeV dark matter or dark photon parameter space.


\begin{acknowledgments}
This work was supported by the Strategic Transformative Research Program at Texas A\&M University (R.M. and M.N.). R.M. acknowledges DOE support through DE-SC0017859 and DE-SC0018981 awards that were instrumental in providing equipment and facilities to carry out this experiment. M.N. is also supported by funding from the Welch foundation (A-1880). T.M. was supported by the World Premier International Research Center Initiative (WPI) MEXT, Japan, and by JSPS KAKENHI grants JP18K13533, JP19H05810, JP20H01896 and JP20H00153. S.R. was supported in part by NSF grant PHY-1638509, the Simons Foundation (award 378243), and the Heising-Simons Foundation grants 2015038 and 2018-0765. Preparation of the Mn12-acetate crystals and the contributions of J.R.L. were supported by the U.S. Department of Energy, Office of Science, Basic Energy Sciences under award DE-SC0025176. All data to be made publicly available within 30 days of the publication of this manuscript \cite{data_drive}.
\end{acknowledgments}

\bibliography{mybibfile}

\end{document}